1# Approximate Optimal Filter for Linear Gaussian Time-invariant Systems

Kaiming Tang, Shengbo Eben Li, Yuming Yin, Yang Guan, Jingliang Duan, Wenhan Cao, Jie Li*Abstract*—State estimation is critical to control systems, especially when the states cannot be directly measured. This paper presents an approximate optimal filter, which enables to use policy iteration technique to obtain the steady-state gain in linear Gaussian time-invariant systems. This design transforms the optimal filtering problem with minimum mean square error into an optimal control problem, called Approximate Optimal Filtering (AOF) problem. The equivalence holds given certain conditions about initial state distributions and policy formats, in which the system state is the estimation error, control input is the filter gain, and control objective function is the accumulated estimation error. We present a policy iteration algorithm to solve the AOF problem in steady-state. A classic vehicle state estimation problem finally evaluates the approximate filter. The results show that the policy converges to the steady-state Kalman gain, and its accuracy is within 2 %.

*Index Terms*—Optimal Estimation, Kalman filter, Reinforcement learning, Dynamic programming.## I. INTRODUCTION

STATE estimation or filtering is a fundamental problem in engineering fields, including communication networks, signal processing, and control systems. A milestone in filter theory is the success of Kalman filter in space projects in the 1960s [1]-[3]. For a time-invariant linear system with mild conditions, the Kalman gain can converge to steady-state Kalman gain $K_\infty$ which can be calculated offline. Steady-state Kalman filter directly applies the steady-state Kalman gain in order to improve the online calculation efficiency. Various steady-state Kalman filter applications such as motor control and neural interfaces system are shown in [4][5]. The computation of steady-state Kalman gain in estimation problems usually requires the solution of a discrete-time algebraic Riccati equation. The solving methods of discrete-time algebraic Riccati equation, such as Schur vector method [6], symplectic SR methods [7][8], doubling method [9], generally require time complexity of $O(n^3)$ and space complexity of $O(n^2)$, which are not suitable for high dimensional systems.

Reinforcement learning [10] is a potential alternative for high-dimensional problems, especially for large-scale optimal control and decision-making problems [11][12]. It has been applied and achieved good performance for various challenging domains, such as Atari games, Go, and robotic control [13]-[16]. Policy iteration is a critical iterative framework widely used in reinforcement learning. There are two revolving iteration procedures for policy iteration: 1) policy evaluation (PEV), which computes the value function for a fixed policy, and 2) policy improvement (PIM), which improves the policy by selecting the actions that maximize the value function computed in the previous step. Policy iteration is able to handle high-dimensional problems when it employs high capacity approximate functions such as neural networks as its policy and value functions [14].

In this paper, we propose the Approximate Optimal Filter (AOF) framework for the purpose of solving the optimal filter gain using reinforcement learning. The AOF transforms the optimal filtering (OF) problem with minimum mean square error into an optimal control problem called the AOF problem, which could be solved by policy iteration. We prove that this transformation does not affect the OF problem's optimality under certain conditions about initial state distributions and policy formats. Subsequently, a policy iteration algorithm under the AOF framework is proposed to solve the OF problem in steady-state, whose approximate solution converges to the optimal one in a vehicle sideslip angle estimation problem. The contributions are: 1) Propose an alternative framework for solving the optimal filter gain in linear Gaussian systems by reinforcement learning, avoiding the problems when solving the discrete-time algebraic Riccati equation in high dimensional systems; 2) Prove the equivalence between the OF and the AOF in linear Gaussian systems, providing the theoretical basis for the problem transformation.

The rest of the paper is organized as follows. In Section II, we describe the OF problem in linear Gaussian system. In Section III, we introduce the AOF framework. In Section IV, we discuss the equivalence between the AOF and the OF. In Section V, a policy iteration algorithm under the AOF framework is described to solve the optimal filter gain. Section VI provides a simulation example to verify our method. Section VII concludes this study.

## II. PROBLEM DESCRIPTION

### A. Linear Gaussian system

The study considers a linear time-invariant Markov model:

$$\begin{cases} x_{t+1} = Ax_t + Bu_t + \xi_t \\ y_t = Cx_t + Du_t + \zeta_t \end{cases} \quad (1)$$

where $t$ is the discrete time, $x_t \in \mathbb{R}^n$ is the state, $u_t \in \mathbb{R}^m$ is the input, $y_t \in \mathbb{R}^r$ is the measurement, $\xi_t \in \mathbb{R}^n$ is the process noise, and $\zeta_t \in \mathbb{R}^r$ is the measurement noise. The system matrices $A, B, C, D$ are known with compatible dimensions.

The assumptions about process noise $\xi_t$, measurement noise $\zeta_t$, and initial state $x_0$ are listed: (1) $\xi_t$ and $\zeta_t$ are individually zero mean, Gaussian white noise with known covariance; (2) $\xi_t$ and $\zeta_t$ are independent, and independent of the initial state $x_0$; and (3) initial state $x_0 \sim \mathcal{N}(\bar{x}_0, \sigma_0^2)$, where $\bar{x}_0$ and $\sigma_0^2$ is the mean and covariance of initial state, respectively. The assumptions can be formulated as

$$\mathbb{E}\{\xi_k\} = 0, \mathbb{E}\{\zeta_k\} = 0, \mathbb{E}\{\xi_k \xi_l^T\} = Q\delta_{k,l}$$
$$, \mathbb{E}\{\zeta_k \zeta_l^T\} = R\delta_{k,l}, \mathbb{E}\{\xi_k \zeta_l^T\} = 0 \quad (2)$$

where $\delta_{k,l}$ is Kronecker delta for all $k$ and $l$, $Q$ is the covariance of process noise, and $R$ is the covariance of measurement noise.

### B. Optimal criterion

The optimal criterion for approximated optimal estimator is to minimize the mean-square-error (MSE) of the estimate and true state in each time step, given the history information $Y$:

$$\min \mathbb{E}\{\|x_t - \hat{x}_t\|_2^2 | Y\} \quad (3)$$

where $\hat{x}_t \in \mathbb{R}^n$ is the estimate and $Y$ represents the history information $(\hat{x}_0, y_{0:t})$ up to time $t$. Note that $\hat{x}_0$ is the initial estimate and $y_{0:t}$ denotes measurement from time step $0$ to $t$. In this study, the optimal estimator is assumed to have a linear structure like

$$\hat{x}_t = A\hat{x}_{t-1} + Bu_{t-1} + L_t[y_t - C(A\hat{x}_{t-1} + Bu_{t-1}) - Du_t] \quad (4)$$

where $L_t \in \mathbb{R}^{n \times r}$ is the filter gain.

The optimal estimation problem seeks to find an optimal filter gain $L_t^* \in \mathbb{R}^{n \times r}$ that can minimize MSE in (3). Let us define the estimation error of each time step as

$$e_t \stackrel{\text{def}}{=} x_t - \hat{x}_t$$

Now the dynamics of the estimation error becomes

$$e_t = (I - L_t C)(Ae_{t-1} + \xi_{t-1}) - L_t \zeta_t \quad (5)$$

For linear systems, the optimal state estimate in the previous step has been proved to contain all the previous information. Therefore, we can transform (3) into one-step optimal criterion like

$$\min_{L_t} \mathbb{E}_{e_t}\{e_t^T e_t\} \quad (6)$$

The conditional expectation is transformed into an unconditional expectation by canceling the information of the linear system model.

### III. APPROXIMATE OPTIMAL FILTER

In this design, we try to transform the MSE estimation into an optimal control problem by generalizing one-step optimal criterion into infinite horizon optimal criterion. The equivalent control problem is then solved by policy iteration technique to obtain the optimal filter gain. Here, we consider the definition of an infinite-horizon Markov decision process (MDP), where the $s \in \mathcal{S}$ is the state, and $a \in \mathcal{A}$ is the action, $s' = f(s, a)$ is the transition dynamics of estimation error. A reward $r$ is given on each state transition.

### A. Approximate Optimal filtering problem formulation

The formulation of approximate optimal filtering (AOF) problem from optimal filtering (OF) is described as follows. The estimation error $e$ is considered as the state $s$, and the filter gain $L$ is regarded as action $a$. Therefore, the environment is the error dynamics, described as

$$s_{i+1} = f(s_i, a_i)$$
$$= (I - a_i C)(As_i + \xi_i) - a_i \zeta_{i+1} \quad (7)$$

where the subscript $i$ is the virtual time step. A deterministic policy $\pi(s)$ is a mapping from state space $\mathcal{S}$ to action space $\mathcal{A}$, i.e.

$$a_i = \pi(s_i)$$

The approximate optimal control problem aims to minimize the discounted accumulative return of reward signals

$$J(\pi) = \mathbb{E}_{s_0 \sim d_0, s_{i+1} \sim f}\left\{\sum_{i=0}^{\infty} \gamma^i r_{i+1}\right\}$$
$$= \mathbb{E}_\pi\left\{\sum_{i=0}^{\infty} \gamma^i r_{i+1}\right\} \quad (8)$$

where $d_0$ is the initial state distribution, $\gamma$ is the discount factor, and $r$ is the reward signal from one-step mean-square-error:

$$r(s, a) = -s^T s = -\text{tr}[ss^T] \quad (9)$$

We use $\mathbb{E}_\pi\{\cdot\}$ to denote the expectation of a random variable relative to the trajectory distribution induced by both policy $\pi$ and initial state $s_0$. The discounting factor is in the range of $0 \leq \gamma < 1$. While $\gamma = 0$, the approximate estimator is "myopic", which only maximizes expected immediate rewards, otherwise the estimator is the expected future accumulated reward.

The state-value function $v^\pi(s)$ is defined as the expected future accumulate reward from the state $s$ induced by policy $\pi$. The state-value function is shown as:

$$v^\pi(s) = \mathbb{E}_\pi\left\{\sum_{i=0}^{\infty} \gamma^i r_{i+1} \bigg| s_0 = s\right\} \quad (10)$$

The action-value $q^\pi(s, a): (\mathcal{S}, \mathcal{A}) \to \mathbb{R}^-$ is the expected future accumulated reward induced by policy $\pi$ after conducting action $a$ from the state $s$, as shown in (11).

$$q^\pi(s, a) = \mathbb{E}_\pi\left\{\sum_{i=0}^{\infty} \gamma^i r_{i+1} \bigg| s_0 = s, a_0 = a\right\} \quad (11)$$

### B. Approximate Optimal Filtering Problem

Combining all the elements in the previous Section, we

summarize an AOF problem as:

$$\max_{\pi} J(\pi) = \mathbb{E}_{\pi}\left\{-\sum_{i=0}^{\infty} \gamma^i s_{i+1}^T s_{i+1}\right\} \quad (12)$$

s.t.

$$s_{i+1} = (I - \pi(s_i)C)(As_i + \xi_i) - a_i\zeta_{i+1}$$

It is easy to see that the objective function in (12) is equal to $\mathbb{E}_{s_0}\{v^\pi(s_0)\}$ and $\mathbb{E}_{s_0}\{q^\pi(s_0, \pi(s_0))\}$. The optimal policy $\pi^*$, i.e., optimal filter gain, is

$$\pi^* = \operatorname*{argmax}_{\pi} v^\pi(s) \quad (13)$$

which has a maximum state-value.

AOF framework formulates an optimal control problem from OF problem. AOF problem could be solved by reinforcement learning algorithms. In Section IV, we will formally prove that both the Kalman gain and the steady-state Kalman gain are the optimal solution of the AOF problem in particular cases, i.e., the AOF problem is equivalent to the OF problem given certain conditions.

IV. EQUIVALENCE OF OPTIMAL FILTERING PROBLEM AND APPROXIMATE OPTIMAL FILTER PROBLEM

This Section will prove that transforming the optimal filter problem into an infinite horizon optimal control problem does not change the optimal action.

A. OF problem in linear Gaussian system

1) The optimal solution of the OF problem:
   The following Lemma gives the form of Kalman gain, which is the optimal solution of the OF problem.
   *Lemma 1*: Kalman filter's recursion
   Kalman filter is often *to write recursion in two steps. In what follows, $\hat{\Sigma}_{n|m}$ and $\hat{x}_{n|m}$ denotes using information up to time $m$ to estimation error covariance and state in time $n$. For $t = 0$, initialize predict error covariance $\hat{\Sigma}_{0|-1} = \sigma_0^2$ and $\hat{x}_{0|-1} = \bar{x}$. For $t \geq 1$, recurse in the following two steps:
   *Time update:*
   Predict state estimate: $\hat{x}_{t|t-1} = A\hat{x}_{t-1|t-1} + Bu_{t-1}$
   Predict error covariance:

$$\hat{\Sigma}_{t|t-1} = A\hat{\Sigma}_{t-1|t-1}A^T + EQE^T \quad (14)$$

   *Measurement update:*
   The Optimal Kalman gain:

$$K_t = \hat{\Sigma}_{t|t-1}C^T\left(C\hat{\Sigma}_{t|t-1}C^T + R\right)^{-1} \quad (15)$$

   Update state estimate: $\hat{x}_{t|t} = \hat{x}_{t|t-1} + K_t(y_t - C\hat{x}_{t|t-1})$
   Update error covariance: $\hat{\Sigma}_{t|t} = A\hat{\Sigma}_{t|t-1}A^T + EQE^T$
   For further detail, please refer to [3].

2) The optimal solution in AOF:
   To discuss the equivalence of AOF problem and OF problem in linear Gaussian system, we start from the optimal criterion defined in (3). If all previous estimates are optimal, we only need to consider one step in linear Gaussian case, the one-step optimal criterion (6) is applied. The optimization problem is:

$$\min_{L_t} J = \mathbb{E}_{x_t}\{(x_t - \hat{x}_t)^T(x_t - \hat{x}_t)|\hat{x}_{t-1}^*, y_t$$
$$= Cx_t + Du_t + \zeta_t\}$$
$$= \mathbb{E}_{e_t}\{e_t^T e_t\} \quad (16)$$

s.t. $e_t = (I - L_t C)(Ae_{t-1}^* + \xi_{t-1}) - L_t \zeta_t$

$$e_{t-1}^* \sim d_{e^*, t-1}$$

where $e_{t-1}^*$ denotes the previous optimal estimation error, which is a random variable that follows a distribution $d_{e^*, t-1}$.

*Proposition 1:* OF problem in time-invariant linear Gaussian system (16) could be solved by AOF framework (12) by setting the initial state distribution $d_0 = d_{e^*, t-1}$. The Kalman gain is the optimal action of action value function weighted by distribution $d_{e^*, t-1}$, i.e.,

$$K_t = \operatorname*{argmax}_{L_t} \mathbb{E}_{s_0}\{q^{\pi^*}(s_0, L_t)\}$$
$$s_0 \sim d_{e^*, t-1} \quad (17)$$

*Proof:* see Appendix B, **Proposition 1**.

B. OF problem in steady-state in linear Gaussian system

1) The optimal solution of the OF problem:
   *Lemma 2*: Steady-state Kalman gain
   The steady-state Kalman gain is pre-calculable if the following statements are fulfilled: Let $Q = E_1 E_1^T$. If $(A, C^T)$ is completely detectable and $(A, EE_1)$ is completely stabilizable, the predicted error covariance matrix $\hat{\Sigma}_{t|t-1}$ would converge to $\hat{\Sigma}$ under Kalman filter recursion. Thus, the steady-state Kalman gain $K_\infty$ could be calculated before any observation is made. $\hat{\Sigma}$ satisfy discrete-time algebraic Riccati equation such that:

$$\hat{\Sigma} = A\hat{\Sigma}A^T - A\hat{\Sigma}C^T(C\hat{\Sigma}C^T + R)^{-1}C\hat{\Sigma}A^T + EQE^T \quad (18)$$

Steady-state Kalman gain $K_\infty$ is:

$$K_\infty = A\hat{\Sigma}C^T(C\hat{\Sigma}C^T + R)^{-1} \quad (19)$$

Steady-state Kalman filter output estimate by:

$$\hat{x}_t = A\hat{x}_{t-1} + Bu_{t-1} + K_\infty(y_t - C\hat{x}_{t|t-1}) \quad (20)$$

The Kalman gain converges when the estimation error becomes a stationary distribution. In other words, the estimation error is steady under a time-invariant gain.

2) The optimal solution in AOF:
   *Proposition 2:* The AOF framework could solve the OF problem in steady-state in linear Gaussian system by setting AOF's initial state as steady-state distribution and using a state-independent time-invariant matrix as the policy directly.

*Proof:* see Appendix B, **Proposition 2,** which *formulates a particular AOF problem:*

$$\max_{\pi} J(\pi) = \mathbb{E}_{\pi}\left\{\sum_{i=0}^{\infty} \gamma^i r_{i+1}\right\}, r_i = -s_i^T s_i \quad (21)$$

$$\text{s.t.} \; s_{i+1} = (I - a_i C)(A s_i + \xi_i) - a_i \zeta_{i+1}$$

$$s_0 \sim d_{e^*, \text{steady}}$$

Note that (21) is a particular case of AOF problem (12) when the initial state is steady. If we restrict our policy to a state-independent time-invariant matrix, i.e., $\pi = L$. Then the optimal policy $\pi^*$ is steady-state Kalman gain. We will verify this result by numerical experiments.

In this Section, we have proved that no matter what initial state distribution $d_0$ is, the optimal action of AOF is given by Kalman filter's recursion. The steady-state Kalman gain is optimal action while the initial state is steady in AOF.

## V. PRACTICAL ALGORITHM

This Section demonstrates a policy iteration algorithm to solve the OF problem in steady-state, one of the applications in AOF framework.

To find an optimal policy in (21), we adopt a policy iteration algorithm called Adaptive Dynamic Programming (ADP) [17]. The flow chart of ADP is shown in Fig. 1. ADP is implemented as an actor-critic architecture involving a critic parameterized function for value function approximation and an actor parameterized function for policy approximation. ADP optimizes the trajectory starting from $s_0$. $k$ is the iteration of ADP.

To solve the optimal estimation problem in steady-state, the environment model is (7) and an initial state sampler is applied. The policy $\pi$ is parameterized as:

$$\pi(s_i; \theta) = a \quad (22)$$

where $\theta$ are the elements of policy $\pi \in \mathbb{R}^{n \times r}$. We approximate state-value as $V(s; w)$ which is a mapping from $\mathcal{S} \to \mathbb{R}^-$ and $w$ is the parameter.

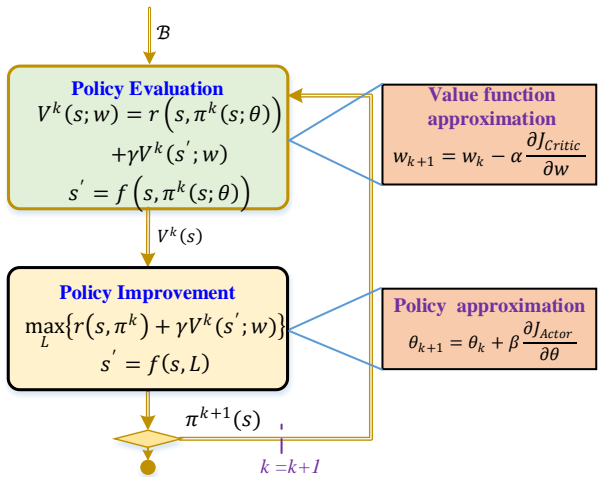

Fig. 1. The flow chart of ADP algorithm.

### A. Sample initial state:

As described in the previous Section, we aim to solve optimal steady-state estimation via solving a particular AOF problem which sets the initial state $s_0 = e_{t-1}^*$ and its distribution $d_0 = d_{e^*, \text{steady}}$. We sample a batch of estimation errors in $d_{e, \text{steady}}$.

$$\mathcal{B} = \{e^{(0)}, e^{(1)}, \dots, e^{(j)}, \dots, e^{(M-1)}\} \quad (23)$$

where $M$ is the batch size.

### B. Policy evaluation (PEV):

In PEV step, we first define the loss function of Critic as:

$$J_{\text{critic}} = \mathbb{E}_{s_i \sim \mathcal{B}}\left\{\frac{1}{2}\big(r_{i+1} + \gamma V(s_{i+1}; w) - V(s_i; w)\big)^2\right\} \quad (24)$$

In (24), we consider the initial state $s_i$ follows the distribution of $\mathcal{B}$. We use the environment model (7) to acquire the next state $s_{i+1}$ and reward $r_{i+1}$ given action $\pi(s_i; \theta)$. We use semi-gradient to optimize (24)

$$\frac{\partial J_{\text{critic}}}{\partial w} = \mathbb{E}_{s_i \sim \mathcal{B}}\left\{\big(r_{i+1} + \gamma V(s_{i+1}; w) - V(s_i; w)\big)\left(-\frac{\partial V(s_i; w)}{\partial w}\right)\right\} \quad (25)$$

Update Critic by gradient descend method.

$$w_{k+1} = w_k - \alpha \frac{\partial J_{\text{critic}}}{\partial w} \quad (26)$$

where $\alpha$ is the learning rate of the Critic.

### C. Policy improvement (PIM):

In PIM step, we define Actor whose loss function is:

$$J_{\text{actor}} = \mathbb{E}_{s_i \sim \mathcal{B}}\{r_{i+1} + \gamma V(s_{i+1}; w)\} \quad (27)$$

where the next state $s_{i+1}$ and reward $r_{i+1}$ for every initial state from PEV step.

The gradient of Actor's loss function is

$$\frac{\partial J_{\text{actor}}}{\partial \theta} = \frac{\partial \mathbb{E}_{s_i \sim \mathcal{B}}\{r_{i+1} + \gamma V(s_{i+1}; w)\}}{\partial \theta} \quad (28)$$

Update Actor by gradient ascend method.

$$\theta_{k+1} = \theta_k + \beta \frac{\partial J_{\text{actor}}}{\partial \theta} \quad (29)$$

where $\beta$ is the learning rate of the Actor.

### D. Algorithm:

The pseudo-code for adaptive dynamic programming is shown as follows:

---

**Algorithm 1** Adaptive Dynamic Programming

Initialize $\theta_0, w_0$.
**Repeat**
  **Sample initial state** (23)
  **PEV step:**
  Calculate Critic loss (24) and its gradient (25)
  Update value function using (26)
  **PIM step:**
  Calculate Actor loss (27) and its gradient (28)
  Update policy using (29)
**until** Convergence

---

## VI. NUMERICAL EXPERIMENT

We apply the algorithm described in Section V.D to solve the AOF problem in steady-state, one particular case in the AOF framework, and compare the obtained policy to steady-state Kalman gain $K_\infty$. We consider a typical vehicle sideslip angle estimation problem with noisy lateral acceleration and yaw rate measurements.

### A. Training environment

In this experiment, we employ a 2-degrees of freedom (DOF) bicycle model. The process model and measurement model are:

$$\begin{cases} \dot{x} = \begin{bmatrix} \dot{\beta}_{\text{SideSlip}} \\ \dot{\omega}_r \end{bmatrix} = Ax + Bu \\ y = \begin{bmatrix} a_y \\ \omega_r \end{bmatrix} = Cx + Du \end{cases} \quad (30)$$

where

$$A = \begin{bmatrix} \frac{(C_f + C_r)}{mv_{\text{long}}} & \frac{(aC_f - bC_r)}{mv_{\text{long}}^2} - 1 \\ \frac{(aC_f - bC_r)}{I_{zz}} & \frac{(a^2C_f + b^2C_r)}{v_{\text{long}}I_{zz}} \end{bmatrix}, B = \begin{bmatrix} -\frac{C_f}{mv_{\text{long}}} \\ -\frac{aC_f}{I_{zz}} \end{bmatrix}$$

$$C = \begin{bmatrix} \frac{(C_f + C_r)}{m} & \frac{(aC_f - bC_r)}{mv_{\text{long}}} \\ 0 & 1 \end{bmatrix}, D = \begin{bmatrix} -\frac{C_f}{m} \\ 0 \end{bmatrix}$$

In (30), $x = [\beta_{\text{SideSlip}} \quad \omega_r]^T$ is state, $\beta_{\text{SideSlip}}$ is sideslip angle, and $\omega_r$ is the yaw rate. Control input is $u = \delta$, where $\delta$ is the front-wheel angle. Output $y = [a_y \quad \omega_r]^T$ is lateral acceleration and yaw rate. Other coefficients are shown in Table 1. The discretized model with sample time 0.01s is applied.

TABLE 1
EXPERIMENT PARAMETER

| | Physical meaning | Value |
|---|---|---|
| $m$ | Mass of the vehicle | 1500 kg |
| $v_{\text{long}}$ | Longitudinal speed | 20 m/s |
| $a$ | Distance from the front-wheel axis to c.g. | 1.14 m |
| $b$ | Distance from rear-wheel axis to c.g. | 1.4 m |
| $C_f$ | Front-wheel cornering stiffness | -44000×2 N/rad |
| $C_r$ | Rear-wheel cornering stiffness | -47000×2 N/rad |
| $I_{zz}$ | Moment of inertia | 2420 kg/m² |
| $\sigma_{\text{SideSlope}}$ | std. of force caused by side slope | 122.625 N |
| $\sigma_{\text{SideWind}}$ | std. of force caused by sidewind | 100 N |
| $\sigma_{\text{LateralAcc}}$ | Lateral acceleration measurement noise std. | 0.05886 m/s² |
| $\sigma_{\text{YawRate}}$ | Yaw rate measurement noise std. | 0.0005814 rad/s² |
| $l_{\text{arm}}$ | Equivalent moment arm of sidewind | -0.13 m |

After the discretization, we consider a discrete time stochastic system based on (30). The process noise $\xi_1$ and $\xi_2$ produced by side-slope and sidewind with noise input matrix $E$ and $l_{\text{arm}}$ is the equivalent moment arm.

$$\xi_t = \begin{bmatrix} \xi_{1,t} \\ \xi_{2,t} \end{bmatrix}, E = \begin{bmatrix} \frac{\Delta t}{mv_{\text{long}}} & \frac{\Delta t}{mv_{\text{long}}} \\ 0 & \frac{l_{\text{arm}}\Delta t}{I_{zz}} \end{bmatrix} \quad (31)$$

$\xi_{1,t} \sim \mathcal{N}(0, \sigma_{\text{SideSlope}}), \xi_{2,t} \sim \mathcal{N}(0, \sigma_{\text{SideWind}})$

In this paper, we build measurement noise from the Bosch SMI700 datasheet. $\zeta_1$ and $\zeta_2$ are measurement noises, where

$$\zeta_t = \begin{bmatrix} \zeta_{1,t} \\ \zeta_{2,t} \end{bmatrix} \quad (32)$$

$\zeta_{1,t} \sim \mathcal{N}(0, \sigma_{\text{LateralAcc}}), \zeta_{2,t} \sim \mathcal{N}(0, \sigma_{\text{YawRate}})$

For further details, please refer to APPENDIX A.

### B. Performance indicator

To eliminate randomness in the experiment result, we average 10000 trajectories. Each trajectory has 1000 steps and starts from different initial states. We study transient and steady-state performance by dividing trajectories into transient and steady-state. The critical time $\tilde{t}$ is where the transient-state becomes steady, which is indicated by the log of MSE. We apply control $u(t)$ while calculating loss:

$$u(t) = \frac{7\pi}{1800}\left[\sin\left(\frac{1}{3\pi t}\right) + \sin\left(\frac{1}{10\pi t}\right) + \sin\left(\frac{1}{20\pi t}\right)\right] \quad (33)$$

In our case $\tilde{t} = 195$, dividing transient and steady-state as shown in the red dash line in Fig. 2. The average loss in the transient-state is defined as:

$$Loss_{\text{tran}} = \frac{1}{N} \sum_{\text{trajectory}} \frac{\sum_1^{\tilde{t}}(x_t - \hat{x}_t)^T(x_t - \hat{x}_t)}{\tilde{t}} \quad (34)$$

where $\tilde{t}$ is the critical time. The trajectory after critical time is defined as steady-state. The average steady-state loss is:

$$Loss_{\text{ss}} = \frac{1}{N} \sum_{\text{trajectory}} \frac{\sum_{\tilde{t}+1}^{T_{test}}(x_t - \hat{x}_t)^T(x_t - \hat{x}_t)}{T_{test} - \tilde{t}} \quad (35)$$

The average loss of full trajectory is:

$$Loss_{full} = \frac{1}{N} \sum_{trajectory} \frac{\sum_1^{T_{test}}(x_t - \hat{x}_t)^T(x_t - \hat{x}_t)}{T_{test}} \quad (36)$$

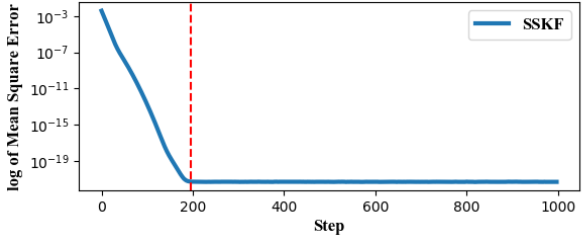

Fig. 2. The log of mean square error of steady-state Kalman Filter

TABLE 2
TESTING PARAMETER

| | Meaning | Value |
|---|---|---|
| $\tilde{t}$ | Critical time | 195 steps |
| $N$ | Trajectory number in the test | 10000 steps |
| $T_{test}$ | Episode-length time in test | 1000 steps |

To compare policy $\pi$ and $K_\infty$, we define the following performance indexes:

The difference of elements in the gain matrix is defined by:

$$\mathcal{D}_{i,j} = (\pi_{i,j} - K_{\infty\ i,j}) \quad (37)$$

The accuracy is defined by:

$$\mathcal{E}_{i,j} = \frac{\mathcal{D}_{i,j}}{|K_{\infty,max}|} \times 100\% \quad (38)$$

where $\#_{i,j}$ indicates the element in row $i$, column $j$ of matrix $\#$ and $|\cdot|$ denotes the absolute value. $K_{\infty,max}$ is the largest element in steady-state Kalman gain $K_\infty$. The analytical $K_\infty$ is calculated from (19).

### C. Solving by AOF

Since the true state value function $v^\pi(s): \mathcal{S} \to \mathbb{R}^-$ is a mapping from state space to negative real number. We set the approximate state-value function $V(s; w)$ as a negative quadratic function of $s$:

$$V(s; w) = -s^T w s \quad (39)$$

where $w$ is a positive-symmetric matrix updated in PEV, and $w$ is initialized as an identity matrix.

$$w_0 = \begin{bmatrix} 1 & 0 \\ 0 & 1 \end{bmatrix} \quad (40)$$

Meanwhile, a time-invariant gain is utilized:

$$\pi(s_i; \theta) = \begin{bmatrix} \theta_{11} & \theta_{12} \\ \theta_{21} & \theta_{22} \end{bmatrix} = a_i \quad (41)$$

and it is initialized as a zero matrix.

We sample the initial state $\mathcal{B}$ from running the environment model given initial estimation error $e_0 = [E_1 \ E_2]^T$.

$$E_1 \sim \mathcal{U}\left[-\frac{5}{180}\pi \ \ \frac{5}{180}\pi\right]$$
$$E_2 \sim \mathcal{U}\left[-\frac{10}{180}\pi \ \ \frac{10}{180}\pi\right] \quad (42)$$

We set the batch size as 256, the learning rate of Actor is 0.003, and the learning rate of Critic is 0.01. Adam method [18] is implemented to update the parameters of Critic and Actor. The discount factor is 0.99.

As shown in Fig. 3, the difference of the four elements in the gain matrix converge to 0, which means the policy $\pi$ converge to steady-state Kalman gain $K_\infty$. Elements in learned policy $\pi^{lr}$ are respectively shown in Table 3. The result shows that we could obtain an optimal filter gain via AOF problem formulation and reinforcement learning algorithm.

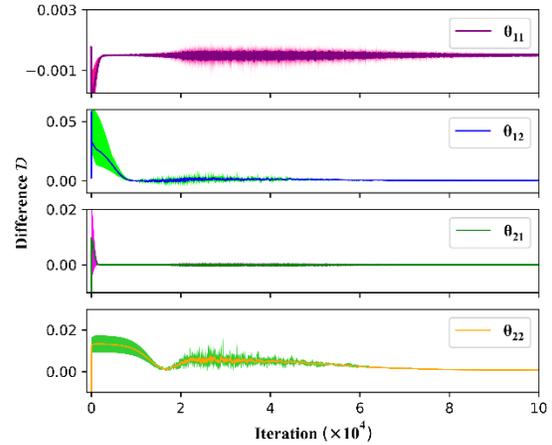

Fig. 3. The differences of the four gain elements during training (averaged over 10 runs)

TABLE 3
COMPARISON OF FILTER GAIN

| | $\theta_{11}$ | $\theta_{12}$ | $\theta_{21}$ | $\theta_{22}$ |
|---|---|---|---|---|
| $K_\infty$ | -5.31e-4 | -2.31e-3 | 3.25e-5 | 5.07e-2 |
| $\pi^{lr}$ | -5.32e-4 | -2.16e-3 | 3.26e-5 | 5.12e-2 |
| $\mathcal{D}$ | -2.50e-7 | 1.54e-4 | 1.32e-7 | 4.66e-4 |
| $\mathcal{E}(\%)$ | -5e-4 | 3.03e-1 | 3e-4 | 9.17e-1 |

### D. Effect of the discount factor

In general, the choice of discount factor affects learned policy. To analyze the effect of different discount factors, we compare different policies learned from different discount factors. Settings are like those of the previous. We sample the initial state $\mathcal{B}$ from running an environment model given the same initial estimation error. The initial estimation error of the sideslip angle is 5 degree and the yaw rate is 10 deg/s:

$$e_0 = \left[\frac{\pi}{36} \ \ \frac{\pi}{18}\right]^T \quad (43)$$

The discount factors are

$$\gamma = \begin{Bmatrix} 0.01 \\ 0.25 \\ 0.5 \\ 0.75 \\ 0.99 \end{Bmatrix}$$

After averaged over 10 runs for each discount factor, 4 elements in the learned policy $\pi^{lr}$ are close to the optimal solution of steady-state Kalman gain $K_\infty$. Filter gains learned from different discount factors are shown in Table 4 and their corresponding performances are shown in Table 5. It founds that average loss and policies are similar close to steady-state Kalman gain $K_\infty$. The accuracy of each element is shown in Table 6. In the general reinforcement learning setting, different discount factors lead to different policies where a larger discount factor makes agents more farsighted to the future reward. However, in Table 6, our policies show no significant effect by choosing different discount factors because the initial state's distribution is steady. If our initial state uses a transient-state and steady-state combination, we might not find an optimal policy represented by a simple matrix.

TABLE 4
FILTER GAIN IN DIFFERENT DISCOUNT FACTOR SET

|  | $K_\infty$ | 0.01 | 0.25 | 0.5 | 0.75 | 0.99 |
|---|---|---|---|---|---|---|
| $\theta_{11}$(e-4) | -5.313 | -5.314 | -5.306 | -5.311 | -5.313 | -5.308 |
| $\theta_{12}$(e-3) | -2.309 | -2.121 | -2.152 | -2.138 | -2.144 | -2.125 |
| $\theta_{21}$(e-5) | 3.25 | 3.2 | 3.28 | 3.19 | 3.15 | 3.25 |
| $\theta_{22}$(e-2) | 5.07 | 5.127 | 5.129 | 5.128 | 5.127 | 5.13 |

TABLE 5
AVERAGE PERFORMANCE IN DIFFERENT DISCOUNT FACTOR SET

|  | $K_\infty$ | 0.01 | 0.25 | 0.5 | 0.75 | 0.99 |
|---|---|---|---|---|---|---|
| $Loss_{tran}$ (e-5) | 8.381 | 8.351 | 8.360 | 8.355 | 8.355 | 8.355 |
| $Loss_{ss}$ (e-22) | 4.745 | 4.747 | 4.750 | 4.749 | 4.747 | 4.750 |
| $Loss_{full}$ (e-5) | 1.637 | 1.632 | 1.633 | 1.633 | 1.632 | 1.633 |

TABLE 6
ACCURACY IN DIFFERENT DISCOUNT FACTOR SET

|  | 0.01 | 0.25 | 0.5 | 0.75 | 0.99 |
|---|---|---|---|---|---|
| $\varepsilon_{11}$(%) | -2e-4 | 1.2 e-3 | 4 e-4 | -1 e-4 | 1 e-3 |
| $\varepsilon_{12}$(%) | 3.71e-1 | 3.11 e-1 | 3.37 e-1 | 3.26 e-1 | 3.63 e-1 |
| $\varepsilon_{21}$(%) | -9.8 e-4 | 6.3 e-4 | 1.24 e-3 | -1.99 e-3 | 4 e-5 |
| $\varepsilon_{22}$(%) | 1.0476 | 1.0791 | 1.0635 | 1.0468 | 1.0986 |

## VII. CONCLUSION

We proposed an Approximate Optimal Filter (AOF) framework in order to solve the optimal filter gain via transforming the optimal filtering (OF) problem with minimum MSE into an equivalent optimal control problem called the AOF problem. We proved the equivalence between the AOF problem and the OF problem in particular parameter settings. Specifically, the Kalman gain and the steady-state Kalman gain in a linear Gaussian OF problem are also the solutions of the AOF problem when the initial state distribution and the policy format are properly chosen. Relying on that, we present a policy iteration algorithm to solve the OF problem in steady-state, one of the applications in the proposed AOF framework. We apply the algorithm to a vehicle sideslip angle estimation problem. Results show that the policy converges to the optimal steady-state Kalman gain, verifying the AOF framework's effectiveness.

## APPENDIX A
### PROCESS NOISE MODELING

The process noise $\xi_1$ and $\xi_2$ produced by side-slope and sidewind with noise input matrix $E$ and $l_{arm}$ is the equivalent moment arm.

$$\xi_t = \begin{bmatrix} \xi_{1,t} \\ \xi_{2,t} \end{bmatrix}, E = \begin{bmatrix} \dfrac{\Delta t}{mv_{\text{long}}} & \dfrac{\Delta t}{mv_{\text{long}}} \\ 0 & \dfrac{l_{\text{arm}}\Delta t}{I_{zz}} \end{bmatrix} \quad (44)$$

$\xi_{1,t} \sim \mathcal{N}(0, \sigma_{\text{SideSlope}}), \xi_{2,t} \sim \mathcal{N}(0, \sigma_{\text{SideWind}})$

Assume the road cross slope of 1.5 ~ 2.5%, the upper boundary is 2.5%, so the lateral component of gravity is $w_1 = \pm(1500 \times 9.81 \times 0.025\text{N})$, which is equal to $\pm 367.875\text{N}$. Assuming the range is in $\pm 3\sigma$, the standard deviation of the gaussian distribution is 122.625N. Suppose that the maximum lateral wind is $w_2 = \pm 300\text{N}$, assuming the range is in $\pm 3\sigma$, and the standard deviation of the gaussian distribution is 100 N. Then the variance of side-slope $\sigma_{\text{SideSlope}} = 122.625$ and side-wind $\sigma_{\text{SideWind}} = 100$.

In the torque term, denotes $Len = a + b$ is the length of the vehicle. Because the lateral force are caused by gravity and lateral wind, the yaw direction torque generate by gravity term is not considered in bicycle model, only consider the lateral wind term. Assume that the lateral wind is generated by uniform distribution, the size has been defined by $\sigma_{\text{SideWind}}$, the equivalent moment is:

$$\begin{aligned} M &= \int_{-b}^{a} \frac{w_2}{Len} x \, dx = \frac{w_2}{2Len} x^2 \Big|_{-b}^{a} \\ &= \frac{w_2}{2Len}(a^2 - b^2) \\ &= \frac{w_2}{2L}(a+b)(a-b) \\ &= \frac{w_2}{2}(a-b) = w_2(\frac{1}{2}Len - b) \end{aligned} \quad (45)$$

the equivalent moment arm $l_{\text{arm}}$ is $\frac{1}{2}Len - b$.

## Appendix B
### PROOF OF PROPOSITION 1

**Proposition 1** *OF problem in time-invariant linear Gaussian system (16) could be solved by AOF framework (12) by setting the initial state distribution $d_0 = d_{e^*,t-1}$. The Kalman gain is the optimal action of action value function weighted by distribution $d_{e^*,t-1}$, i.e.,*

$$K_t = \underset{L_t}{\operatorname{argmax}} \mathbb{E}_{s_0}\{q^{\pi^*}(s_0, L_t)\} \quad (46)$$
$$s_0 \sim d_{e^*,t-1}$$

**Proof:**

Since action-value consider infinite horizon. We need to prove the optimal action does not change with considering n steps, n ≥ 1. First, consider only one step objective function $J_1$,

$$J_1 = \mathbb{E}_{s_0,\xi_0,\zeta_1}\{-s_1^T s_1\} = \mathbb{E}_{s_0,\xi_0,\zeta_1}\{\text{tr}[-s_1 s_1^T]\}$$
$$= -\text{tr}[(I - a_0 C)A \mathbb{E}_{s_0}\{s_0 s_0^T\}A^T(I - a_0 C)^T \quad (47)$$
$$+(I - a_0 C)Q(I - a_0 C)^T + a_0 R a_0^T]$$

Denote $\mathbb{E}_{s_0}\{s_0 s_0^T\}$ as $P_0$. The saddle point only happens when $\frac{\partial J_1}{\partial a_0} = 0$ yield

$$\frac{\partial J_1}{\partial a_0} = 2(CAP_0 A^T + CQ)^T \quad (48)$$
$$-2a_0(CAP_0 A^T C^T + CQC^T + R) = 0$$

i.e.

$$a_0^* = (CAP_0 A^T + CQ)^T (CAP_0 A^T C^T + CQC^T + R)^{-1} \quad (49)$$

For any time step $i \geq 1$, denote that

$$P_i(a_{i-1}) = (I - a_{i-1}C)AP_{i-1}A^T(I - a_{i-1}C)^T$$
$$+(I - a_{i-1}C)QA^T(I - a_{i-1}C)^T \quad (50)$$
$$+a_{i-1}R a_{i-1}^T$$

Next, consider the two-step objective function $J_2$,

$$J_2 = \mathbb{E}_{s_0,\xi_0,\zeta_1,s_1,\xi_1,\zeta_2}\{-s_1^T s_1 - \gamma s_2^T s_2\}$$
$$= \mathbb{E}_{s_0,\xi_0,\zeta_1,s_1,\xi_1,\zeta_2}\{-\text{tr}[s_1 s_1^T] - \gamma \text{tr}[s_2 s_2^T]\}$$
$$= -\text{tr}[P_1(a_0) \quad (51)$$
$$+\gamma(I - a_1 C)AP_1(a_0)A^T(I - a_1 C)^T$$
$$+(I - a_1 C)Q(I - a_1 C)^T + a_1 R a_1^T]$$

The saddle point only happens when $\frac{\partial J_2}{\partial a_0} = 0$ and $\frac{\partial J_2}{\partial a_1} = 0$ yield

$$\frac{\partial J_2}{\partial a_0} = 2(CAP_0 A^T + CQ)^T$$
$$-2a_0(CAP_0 A^T C^T + CQC^T + R)$$
$$-\gamma\{A^T(I - a_1 C)^T(I - a_1 C)A \quad (52)$$
$$[-2(CAP_0 A^T + CQ)^T$$
$$+2a_0(CAP_0 A^T C^T + CQC^T + R)]\}^T$$
$$= 0$$

and

$$\frac{\partial J_2}{\partial a_1} = 2\gamma(CAP_1(a_0)A^T + CQ)^T \quad (53)$$

$$-2\gamma a_1(CAP_1(a_0)A^T C^T + CQC^T + R)$$
$$= 0$$

i.e.

$$a_0^* = (CAP_0 A^T + CQ)^T$$
$$(CAP_0 A^T C^T + CQC^T + R)^{-1}$$
$$a_1^* = (CAP_1^* A^T + CQ)^T \quad (54)$$
$$(CAP_1^* A^T C^T + CQC^T + R)^{-1}$$

The optimal $a_0^*$ is same as $J_1$. where $P_1^*$ is $P_1(a_0^*)$. It is proved by mathematical induction that the optimal solution for $n$ steps is given by:

$$a_0^* = (CAP_0 A^T + CQ)^T$$
$$(CAP_0 A^T C^T + CQC^T + R)^{-1}$$
$$\vdots \quad (55)$$
$$a_n^* = (CAP_n^* A^T + CQ)^T$$
$$(CAP_n^* A^T C^T + CQC^T + R)^{-1}$$

where $P_n^* = P_n(a_0^*, a_1^*, \ldots, a_{n-1}^*)$. This solution is the same as Kalman filter's recursion and does not affect by the discount factor $\gamma$.

If $s_0 \sim d_{e^*,t-1}$

$$K_t = \underset{L_t}{\text{argmax}}\, \mathbb{E}_{s_0}\{q^{\pi^*}(s_0, L_t)\}$$
$$= (CAP_{e^*,t-1}A^T + CQ)^T \quad (56)$$
$$(CAP_{e^*,t-1}A^T C^T + CQC^T + R)^{-1}$$

where $P_{e^*,t-1} = \mathbb{E}_{e_{t-1}^*}\{e_{t-1}^* e_{t-1}^{*T}\}$. The statement is proofed immediately.

PROOF OF PROPOSITION 2

***Proposition 2:*** *The AOF framework could solve the OF problem in steady-state in linear Gaussian system by setting AOF's initial state as steady-state distribution and using a state-independent time-invariant matrix as the policy directly。*

***Proof:***

When *t*he terminal time $T$ is sufficiently large, the estimation error distribution becomes stationary distribution $d_{e,\text{steady}}$ under $L_t$ which makes spectral radius $\rho[(I - L_t C)A] = \underset{1 \leq i \leq n}{\max}|\lambda_i| < 1$. Following this, we derive the OF problem in steady-state from (16):

$$\underset{L_t}{\min} J = \mathbb{E}_{e_t}\{e_t^T e_t\} \cong \underset{T \to \infty}{\lim} \frac{1}{T}\sum_{t}^{t+T-1} e_t^T e_t \quad (57)$$
$$\text{s.t. } e_t = (I - L_t C)(Ae_{t-1}^* + \xi_{t-1}) - L_t \zeta_t$$

$$e_{t-1}^* \sim d_{e^*,\text{steady}}$$

where previous optimal estimation error $e_{t-1}^*$ is in steady-state and $d_{e^*,\text{steady}}$ is its distribution. AOF framework could extend to (57) by setting the filter gain is given by policy $\pi$ and the initial state $s_0 = e_{t-1}^*, d_0 = d_{e^*,\text{steady}}$. Optimization object could be extended to average reward or discounted reward by[19]:

$$\lim_{T \to \infty} \frac{1}{T} \sum_{i=0}^{i+T-1} s_{i+1}^T s_{i+1}$$
$$= \lim_{T \to \infty} \frac{1}{T} \sum_{i=0}^{i+T-1} \mathbb{E}_\pi \{s_{i+1}^T s_{i+1}\} \quad (58)$$
$$= \frac{1}{1-\gamma} \sum_{i=0}^{\infty} \mathbb{E}_\pi \{-\gamma^i r_{i+1}\}$$

From (57) and (58), we formulate a particular AOF problem:

$$\max_\pi J(\pi) = \mathbb{E}_\pi \left\{ \sum_{i=0}^{\infty} \gamma^i r_{i+1} \right\}, r_i = -s_i^T s_i$$
$$\text{s.t. } s_{i+1} = (I - a_i C)(A s_i + \xi_i) - a_i \zeta_{i+1} \quad (59)$$
$$s_0 \sim d_{e^*,\text{steady}}$$

which is a particular case of AOF problem (12) when the initial state is steady. If we restrict our policy to a state-independent time-invariant matrix, i.e., $\pi = L$. Then the optimal policy $\pi^*$ is steady-state Kalman gain.